\documentclass[10pt,aps,prd,floatfix,notitlepage,showpacs,nofootinbib,superscriptaddress,onecolumn,showkeys]{revtex4-1}
\usepackage{amssymb,amsmath}

\begin{document}

%\usepackage[T1]{fontenc}
%\usepackage[latin1]{inputenc}
%\usepackage{amsmath, amssymb, amsthm}
%\usepackage[english]{babel}
%\usepackage{indentfirst}
%\usepackage{booktabs}
%\usepackage{array}
%\usepackage{caption,appendix}
%\usepackage{geometry}
%\usepackage{float}
%\usepackage{cancel}
%\usepackage{bbold}
%\usepackage{authblk}
%\numberwithin{equation}{section}
%\begin{document}
\author{Salvatore Capozziello}
\affiliation{Dipartimento di Fisica "E. Pancini", Universit\`a		   di Napoli {}``Federico II'', Compl. Univ. di
		   Monte S. Angelo, Edificio G, Via Cinthia, I-80126, Napoli, Italy, }
		  \affiliation{INFN Sezione  di Napoli, Compl. Univ. di
		   Monte S. Angelo, Edificio G, Via Cinthia, I-80126,  Napoli, Italy,}
 \affiliation{Gran Sasso Science Institute, Via F. Crispi, 7, I-67100, L'Aquila, Italy,}
 \affiliation{ Laboratory for Theoretical Cosmology, Tomsk State University of Control Systems and
Radioelectronics (TUSUR), 634050 Tomsk,  Russia, }
\author{Maurizio Capriolo} 
\affiliation{Dipartimento di Matematica Universit\`a di Salerno, via Giovanni Paolo II, 132, Fisciano, SA I-84084, Italy.} 
\author{Loredana Caso}
\affiliation{Dipartimento di Matematica Universit\`a di Salerno, via Giovanni Paolo II, 132, Fisciano, SA I-84084, Italy.} 
\date{\today}
\title{\textbf{Weak Field Limit and Gravitational Waves in
 $f(T,B)$ Teleparallel Gravity}}

\begin{abstract}
We derive the gravitational waves for $f\left(T, B\right)$ gravity which is  an  extension of  teleparallel gravity  and demonstrate that it is  equivalent to $f(R)$ gravity by  linearized the field equations in the weak field limit approximation.  $f(T,B)$ gravity shows three polarizations: the two standard of general relativity, plus and cross, which are  purely transverse with two-helicity, massless tensor polarization modes, and  an additional massive scalar mode with zero-helicity. The last one is a mix of longitudinal and transverse breathing scalar polarization modes.  The boundary term $B$ excites the extra scalar polarization and the mass of scalar field breaks the symmetry of the TT gauge by adding a new degree of freedom, namely a single mixed scalar polarization.
\end{abstract} 
\date{\today}
\maketitle

\section{Introduction}
Albert Einstein, in 1928, made an attempt to formulate a unified  theory of gravity and electromagnetism by using the geometric notion of teleparallelism introduced a few years before by Cartan. For this purpose,  he relaxed the hypothesis of connection symmetry by Levi Civita and considered a curvature-free connection with torsion, the Weitzenb$\ddot{o}$ck one, and formulated the theory adopting   a tangent space basis that had the property to make the spacetime parallelizable. Then, he used the tetrad $\left\{e_{a}\right\}$  based on the notion of distant or absolute parallelism. This attempt to unify General Relativity (GR) with electromagnetism proved unsuccessful because the components of the electromagnetic field, identified with the additional six components of the tetrad,  could be eliminated by imposing the local Lorentz invariance. However, this alternative formulation, based on geometry modification \cite{MALTEL, CRTG, KSTE}, is equivalent to GR and was named Teleparallel Equivalent General Relativity (TEGR), in the sense that it describes the same physics because it gives  the same field equations of GR. In fact, considering the  Hilbert-Einstein Lagrangian, linear in the Ricci scalar curvature $R$ \cite{APTG, CCLS},
\begin{equation}
L_{HE}\left(g\right)=-\frac{1}{2\kappa^{2}}R\sqrt{-g}\ ,
\end{equation}
with $\kappa^{2}=8\pi G/c^{4}$ and the teleparallel Lagrangian, linear in the torsion scalar  $T$
\begin{equation}
L_{TEGR}\left(e\right)=\frac{e}{2\kappa^{2}}T\ ,
\end{equation}
they differ from each other by a four divergence which is
\begin{equation}
L_{HE}\left(e\right)=L_{TEGR}\left(e\right)+\partial_{\mu}\left(\frac{e}{\kappa^{2}}T^{\rho\mu}_{\phantom{\rho\mu}\rho}\right)\ .
\end{equation}
Several issues of today physics can be addressed by extending  the geometric sector of the Einstein field equations. For example,   $f(R)$ gravity is an extension of GR because  it  the Hilbert-Einstein Lagrangian, linear in the   Ricci curvature scalar  $R$, is extended considering a generic function of it   \cite{CL}. In the same way it is possible to extend the TEGR by considering an analytical function $f (T)$ of the torsion scalar $T$ \cite{FETMG}. 

The   $f(T)$ teleparallel gravity differs from  $f(R)$ gravity  because the former leads to second-order field equations while the latter leads to fourth order field equations in metric formalism.  Furthermore, $f(T)$ gravity is not invariant under local Lorentz transformation if the spin connection is set to zero.  $f\left(T\right)$ gravity can be adopted, for example, to explain the accelerated expansion of the Universe at the present time without the introduction of dark energy (see \cite{CCLS} for a review). If we want to study higher order telepallel theories, equivalent  to those expressed in terms of $R$, we can not limit ourselves to $f(T)$, because it always produces second order dynamical equations. We have to  introduce both boundary term $B=2\nabla_{\mu}\left(T^{\mu}\right)$, depending on the derivatives of the torsion vector $T^{\mu}$ and  terms like $\Box T$, $\Box^{k}T$ in the teleparallel Lagrangian \cite{OS, BC}. We can therefore start from  $f\left(R\right)$ gravity, and find its teleparallel equivalent  after observing that the boundary term is $B=-T-R$  and then restore the  $f\left(T, B\right)$ gravity \cite{camci}. % Indeed $f\left(T, B\right)$-gravity is more general theory of $f\left(R\right)$-gravity or $f\left(T\right)$-gravity because $B$ can be unrelated to $T$, that is, $B$ and $T$ can be considered as two independent fields.
The teleparallel theory of gravity $f(T,B)$ is the teleparallel equivalent of $f(R)$ as the TEGR is the teleparallel equivalent of GR as we will show below by considering the weak field limit and the gravitational wave modes. In the framework of $f(T,B)$ gravity, it is possible to explore the validity of laws of thermodynamics \cite{Bahamonde:2016cul} and derive  energy constraints for   de Sitter (dS), power-law, $\Lambda$CDM and phantom models \cite{Zubair:2018wyy}.  %Otherwise we can generally start directly from a theory of gravity of order $n$, using a telaparallel geometry without the boundary term $B$ but only via the $\Box^{k}T$, necessary to go up with the degree of the theory. Then we can study the field equations associated with the Lagrangian $L_{\Box^{k}T}=e\left(T+a_{0}T^{2}+\sum_{k=1}^{p} a_{k}T\Box^{k}T\right)$.\\

The detection of gravitational waves (GWs)  opened new perspectives in the study of the alternative theories of gravity and, in general, in relativistic astrophysics. In  generic metric theories of gravity, it is possible to show that  the GWs polarizations can give, at maximum of six modes in 4D spacetimes.  More precisely, according to  \cite{ELL,ELLWW}, we have: breathing ($b$), longitudinal ($l$), vector-$x$ ($x$), vector-$y$ ($y$), plus ($+$) and cross ($\times$) modes. 

In order to  study the further GW polarizations,  beyond the two standards plus and cross modes, it is useful  to extend GR to more general theories.  If scalar or vector modes are found, it could mean that  theory of gravitation should be extended beyond GR and  some theoretical models should be excluded.

To this end,  the GW170817 event \cite{GW17_1} set constraints  on viable  gravitational theories. In fact,  the event was the first to   provide constraints on  the speed of electromagnetic  and  gravitational waves.  According to this result, it is possible to  fix  possible masses of  further gravitational  modes  \cite{GW17_2}. 
 This fact is important  to discriminate among concurring   gravitational theories and  some alternatives to GR, including some  scalar-tensor theories like Brans-Dicke gravity, Horava-Lifshitz gravity, and bimetric gravity, seem excluded \cite{GW17_3}.  In particular,  observational constraints on $f(T)$ gravity can be imposed by the combined observation of GW170817 and its electromagnetic counterpart GRB170817A,   as discussed in  \cite{CLSX, NPS}. In these papers,   constraints derived from primordial gravitational waves  are also taken into account.

In summary,  GWs polarizations are a powerful tool to probe theories of gravity. Moreover, by means of the linearized gravitational energy-momentum pseudo-tensor of $f(R)$, $f(T)$ gravity \cite{CCT} or more generally of $f(R, R\Box R, \dots R\Box^{k}R)$ gravity \cite{CCT2}, it  is possible to express the pseudo-tensor in terms of the further modes in order to  test alternative theories of gravity.  
In the framework of teleparallelism,  gravitational waves have   started to be studied recently. These studies led to the interesting possibility to classify teleparallel theories  according to their degrees of freedom\cite{AC,Hohmann1,Hohmann2,Hohmann3}.

In this paper, we  investigate  GWs generated in theories  containing  the torsion scalar $T$ and the boundary term $B$ and show, from this point of view, their equivalence with $f(R)$ gravity.

The layout of the article is as follows: in Sec. \ref{WFL} we obtain the geometrical and physical quantities of interest after  the expansion of tetrads around the flat geometry at first order in the weak field approximation. In Sec. \ref{WFLFTB}, we prove the equivalence between $f(T,B)$ and $f(R)$ theories and then we derive the field equations in presence of matter for  $f\left(T, B \right)$ gravity  in the low energy limit. GWs in vacuum are obtained in  Sec. \ref{GWTP}  and, finally, in Sec. \ref{POLHEL}  both polarization and helicity of GWs are studied by  mean the equation of geodesic deviation and the Newman-Penrose formalism. Conclusions are drawn in Sec. \ref{conc}.

Throughout this work we will use  conventions by Landau and Lifshitz \cite{LL}, that is:
\begin{itemize}
\item[(1)]The metric  signature is $\left(+, -, -, -\right)$\ .
\item[(2)]The Riemann tensor $R^{\rho}_{\phantom{\rho}\lambda\nu\mu}$  for a generic connection $\Gamma$ is defined as 
\begin{equation}
R^{\rho}_{\phantom{\rho}\lambda\nu\mu}=\partial_{\nu}\Gamma^{\rho}_{\phantom{\rho}\lambda\mu}-\partial_{\mu}\Gamma^{\rho}_{\phantom{\rho}\lambda\nu}+\Gamma^{\rho}_{\phantom{\rho}\eta\nu}\Gamma^{\eta}_{\phantom{\eta}\lambda\mu}-\Gamma^{\rho}_{\phantom{\rho}\eta\mu}\Gamma^{\eta}_{\phantom{\eta}\lambda\nu}\ .
\end{equation}
\item[(3)]The Ricci tensor is defined as the contraction $R_{\mu\nu}=R^{\lambda}_{\phantom{\lambda}\mu\lambda\nu}$\ .
\end{itemize}
\section{Weak Field Limit in Teleparallel Gravity}\label{WFL}
 Dynamical variables used in  teleparallelism are the components of  tetrad basis $\left\{e_{a}\right\}$ and  dual basis $\left\{e^{a}\right\}$  which form a local orthonormal basis for the tangent space at each point  $\{x_{\mu}\}$  of the spacetime manifold. The components of the vierbein satisfy relations \cite{CSG, WALD, APIGP, STRAGR, AP}
\begin{equation}
e^{a}_{\phantom{a}\mu}e_{a}^{\phantom{a}\nu}=\delta_{\mu}^{\nu}\ , \qquad \text{and} \qquad e^{a}_{\ \mu}e_{b}^{\ \mu}=\delta_{a}^{b}\ ,
\end{equation}
\begin{equation}
\eta_{ab}=g_{\mu\nu}e_{a}^{\ \mu}e_{b}^{\phantom{b}\nu}\ ,\qquad \text{and}\qquad g_{\mu\nu}=\eta_{ab}e^{a}_{\phantom{a}\mu}e^{b}_{\phantom{b}\nu}\ ,
\end{equation}
where we are going to use the Greek alphabet to denote indices related to spacetime, and the Latin alphabet to denote indices related to the tangent space. The Weitzenb$\ddot{o}$ck  connection is defined as
\begin{equation}\label{Weitzconn}
\tilde{\Gamma}^{\rho}_{\phantom{\rho}\mu\nu}=e_{a}^{\phantom{a}\rho}\partial_{\nu}e^{a}_{\phantom{a}\mu}\ ,
\end{equation}
and its torsion tensor is
\begin{equation}\label{torsiontensor}
T^{\nu}_{\phantom{\nu}\rho\mu}=\tilde{\Gamma}^{\nu}_{\phantom{\nu}\mu\rho}-	\tilde{\Gamma}^{\nu}_{\phantom{\nu}\rho\mu}=e_{a}^{\phantom{a}\nu}\partial_{\rho}e^{a}_{\phantom{a}\mu}-e_{a}^{\phantom{a}\nu}\partial_{\mu}e^{a}_{\phantom{a}\rho}\ .
\end{equation}
Defined the contortion tensor as the connection of Weitzenb$\ddot{o}$ck minus the Levi Civita connection that is
\begin{equation}\label{contortiontensor}
K^{\nu}_{\phantom{\nu} \rho\mu}=\tilde{\Gamma}^{\nu}_{\phantom{\nu} \rho\mu}-\stackrel{\circ}{\Gamma}{\hspace{-0.04in}}^{\nu}_{\phantom{\nu} \rho\mu}=\frac{1}{2}\left(T_{\rho\phantom{\nu} \mu}^{\phantom{\rho}\nu}+T_{\mu\phantom{\mu}\rho}^{\phantom{\mu}\nu}-T^{\nu}_{\phantom{\nu}\rho\mu}\right)\ ,
\end{equation}
and the superpotential tensor $S^{\rho\mu\nu}$ as 
\begin{equation}
S^{\rho\mu\nu}=\frac{1}{2}\left(K^{\mu\nu\rho}-g^{\rho\nu}T^{\sigma\mu}_{\sigma}+g^{\rho\mu}T^{\sigma\nu}_{\sigma}\right)\ ,
\end{equation}
we obtain the scalar torsion $T$
\begin{equation}
T=T_{\rho\mu\nu}S^{\rho\mu\nu}\ ,
\end{equation}
from the contraction of the torsion tensor with the superpotential.
The curvature of the Weitzenb$\ddot{o}$ck connection is $R[\tilde{\Gamma}]=0$, where the Riemann tensor $R^{\rho}_{\phantom{\rho}\lambda\nu\mu}$  for a Weitzenb$\ddot{o}$ck connection is defined as
\begin{equation}
R^{\rho}_{\phantom{\rho}\lambda\nu\mu}=\partial_{\nu}\tilde{\Gamma}^{\rho}_{\phantom{\rho}\lambda\mu}-\partial_{\mu}\tilde{\Gamma}^{\rho}_{\phantom{\rho}\lambda\nu}+\tilde{\Gamma}^{\rho}_{\phantom{\rho}\eta\nu}\tilde{\Gamma}^{\eta}_{\phantom{\eta}\lambda\mu}-\tilde{\Gamma}^{\rho}_{\phantom{\rho}\eta\mu}\tilde{\Gamma}^{\eta}_{\phantom{\eta}\lambda\nu}\ .
\end{equation}
We now express the scalar curvature $R[\stackrel{\circ}\Gamma]$ of the Levi-Civita connection in terms of the scalar torsion $T$ and the vector torsion $T^{\sigma}$, that is  
\begin{equation}
-R[\stackrel{\circ}\Gamma]=T+\frac{2}{e}\partial_{\sigma}\left(eT^{\nu\sigma}_{\phantom{\nu\sigma}\nu}\right)\ ,
\end{equation}
with $T^{\sigma}$ obtained by contracting the first and third torsion tensor index 
\begin{equation}
T^{\sigma}=T^{\nu\sigma}_{\phantom{\nu\sigma}\nu}\ ,
\end{equation}
%\begin{equation}
%-R=T+\frac{2}{e}\partial_{\sigma}\left(eT^{\nu\sigma}_{\phantom{\nu\sigma}\nu}\right)\ ,
%\end{equation}
where $e=det\left(e^{a}_{\phantom{a}\rho}\right)$.
If we indicate the boundary term as
\begin{equation}
B=\frac{2}{e}\partial_{\sigma}\left(eT^{\nu\sigma}_{\phantom{\nu\sigma}\nu}\right)\ ,
\end{equation}
we get the relation\footnote{For the signature of boundary term, see the discussion in \cite{CCT}.} 
\begin{equation}
-R[\stackrel{\circ}\Gamma]=T+B\ .
\end{equation}
We now expand the tetrad field around the flat geometry described by the trivial tetrad $e^{a}_{\phantom{a}\mu}=\delta^{a}_{\phantom{a}\mu}$ as follows
\begin{equation}
e^{a}_{\phantom{a}\mu}=\delta^{a}_{\phantom{a}\mu}+E^{a}_{\phantom{a}\mu}\ ,
\end{equation}
where $\lvert E^{a}_{\phantom{a}\mu} \rvert \ll 1$. 
Thus perturbing the metric tensor $g_{\mu\nu}$ to first order  in $E^{a}_{\phantom{a}\mu}$ we obtain 
\begin{equation}
g_{\mu\nu}=\eta_{\mu\nu}+h_{\mu\nu}+\mathcal{O}\left(h^2\right)=\eta_{\mu\nu}+\eta_{\mu a}E^{a}_{\phantom{a}\nu}+\eta_{\nu a}E^{a}_{\phantom{a}\mu}+\mathcal{O}\left(E^2\right)\ ,
\end{equation}
and so 
\begin{equation}
h_{\mu\nu}=\eta_{\mu a}E^{a}_{\phantom{a}\nu}+\eta_{\nu a}E^{a}_{\phantom{a}\mu}\ .
\end{equation}
The Weitzenb$\ddot{o}$ck connection to first order in $E^{a}_{\phantom{a}\mu}$ becomes
\begin{equation}
\tilde{\Gamma}^{\rho\left(1\right)}_{\phantom{\rho}\mu\nu}=\delta_{a}^{\phantom{a}\rho}\partial_{\nu}E^{a}_{\phantom{a}\mu}\ , .
\end{equation}
The covariant derivative $\nabla_{\mu}$ and the covariant d'Alembert operator $\Box=g^{\mu\nu}\nabla_{\mu}\nabla_{\nu}$ to zero order become
\begin{equation}
\nabla_{\mu}^{\left(0\right)}=\partial_{\mu}\ ,
\end{equation}
\begin{equation}
\Box^{\left(0\right)}=\eta^{\mu\nu}\partial_{\nu}\partial_{\mu}=\partial^{\mu}\partial_{\mu}\ .
\end{equation}
The torsion tensor $T^{\mu}_{\phantom{\mu}\nu\rho}$ and its contraction $T^{\mu\nu}_{\phantom{\mu\nu}\mu}$  can be written as 
\begin{equation}
T^{\mu\left(1\right)}_{\phantom{\mu}\nu\rho}=\delta_{a}^{\phantom{a}\mu}\left(\partial_{\nu}E^{a}_{\phantom{a}\rho}-\partial_{\rho}E^{a}_{\phantom{a}\nu}\right)\ ,
\end{equation}
and 
\begin{equation}
T^{\rho\sigma\left(1\right)}_{\phantom{\rho\sigma}\rho}=\delta_{a}^{\phantom{a}\mu}\eta^{\sigma\nu}\left(\partial_{\nu}E^{a}_{\phantom{a}\mu}-\partial_{\mu}E^{a}_{\phantom{a}\nu}\right)\ .
\end{equation}
We compute the contortion tensor as
\begin{equation}
K^{\rho\left(1\right)}_{\mu\nu}=\eta_{\mu\lambda}\delta^{\phantom{a}\lambda}_{a}\partial^{\rho}E^{a}_{\phantom{a}\nu}-\delta_{a}^{\phantom{a}\rho}\partial_{\mu}E^{a}_{\phantom{a}\nu}\ ,
\end{equation} 
and the superpotential $S_{\rho}^{\phantom{\rho}\mu\nu}$, the scalar torsion $T$ and the boundary term $B$ as 
\begin{align}
2S_{\rho}^{\phantom{\rho}\mu\nu\left(1\right)}=& \delta_{a}^{\phantom{a}\nu}\partial^{\mu}E^{a}_{\phantom{a}\rho}-\delta_{a}^{\phantom{a}\mu}\partial^{\nu}E^{a}_{\phantom{a}\rho}-\delta^{\nu}_{\rho}\left(\delta_{a}^{\phantom{a}\sigma}\partial^{\mu}E^{a}_{\phantom{a}\sigma}-\eta^{\alpha\mu}\delta^{a}_{\phantom{a}\alpha}\partial_{\sigma}E^{\sigma}_{\phantom{\sigma}a}\right)\nonumber\\
&+\delta^{\mu}_{\rho}\left(\delta_{a}^{\phantom{a}\sigma}\partial^{\nu}E^{a}_{\phantom{a}\sigma}-\eta^{\alpha\nu}\delta^{a}_{\phantom{a}\alpha}\partial_{\sigma}E^{\sigma}_{\phantom{\sigma}a}\right)\ ,
\end{align}
\begin{equation}
T^{\left(2\right)}=T^{\mu\nu\rho\left(1\right)}S^{\left(1\right)}_{\mu\nu\rho}\ ,
\end{equation}
\begin{equation}
B^{\left(1\right)}=\left(\frac{2}{e}\partial_{\sigma}\left(e T^{\sigma}\right)\right)^{\left(1\right)}=2\delta_{a}^{\phantom{a}\nu}\left[\Box E^{a}_{\phantom{a}\nu}-\partial^{\mu}\partial_{\nu}E^{a}_{\phantom{a}\mu}\right]\ .
\end{equation}
%where $e=det\left(e^{a}_{\phantom{a}\nu}\right)$ . 
The Ricci curvature $R$, to the first order in $E^{a}_{\phantom{a}\mu}$,  takes the following form
\begin{equation}
R^{\left(1\right)}=-B^{\left(1\right)}\ ,
\end{equation}
that is, the first order boundary term $B^{(1)}$ contributes to the Ricci curvature.
Finally we obtain the useful relation 
\begin{align}\label{DerSupPot}
2\partial_{\nu}S_{\rho}^{\phantom{\rho}\mu\nu(1)}=&\delta_{a}^{\phantom{a}\nu}\partial_{\nu}\partial^{\mu}E^{a}_{\phantom{a}\rho}-\delta_{a}^{\phantom{a}\mu}\Box E^{a}_{\phantom{a}\rho}-\delta_{a}^{\phantom{a}\sigma}\partial_{\rho}\partial^{\mu}E^{a}_{\phantom{a}\sigma} \nonumber\\
&+\eta^{\rho\mu}\delta_{a}^{\phantom{a}\sigma}\partial_{\rho}\partial_{\sigma}E^{a}_{\phantom{a}\rho}+\delta^{\mu}_{\rho}\delta_{a}^{\phantom{a}\sigma}\Box E^{a}_{\phantom{a}\sigma}-\delta^{\mu}_{\rho}\delta_{a}^{\phantom{a}\sigma}\eta^{\alpha\nu}\partial_{\nu}\partial_{\sigma}E^{a}_{\phantom{a}\alpha}\ ,
\end{align}
and we set 
\begin{equation}
E_{\mu\nu}=\eta_{\mu a}E^{a}_{\phantom{a}\nu}\ , \qquad E=\delta_{a}^{\phantom{a}\mu}E^{a}_{\phantom{a}\mu}\ .
\end{equation}
The first order perturbative tetrad $E^{a}_{\phantom{a}\mu}$ is not symmetric because the $f\left(T,B\right)$ gravity is not invariant under a local Lorentz transformation \cite{AC, OP}
\begin{equation}
\eta_{\mu a}E^{a}_{\phantom{a}\nu}\neq \eta_{\nu a}E^{a}_{\phantom{a}\mu}\ ,
\end{equation}
and then, we decompose the perturbation tetrad $E_{\mu\nu}$ into symmetric and antisymmetric parts
\begin{equation}
 E_{\mu\nu}=E_{\left(\mu\nu\right)}+E_{\left[\mu\nu\right]}\ .
\end{equation}  
However, the  antisymmetric  part $E_{\left[\mu\nu\right]}$   have no physical meaning because it is not involved 
into the Lagrangian $L_{f\left(T,B\right)}$ and field equations, depending on the symmetric part $E_{\left(\mu\nu\right)}$ by means of $T$ and $B$.
Hence, we can set to zero the antisymmetric component $E_{\left[\mu\nu\right]}$
\begin{equation}
E_{\left[\mu\nu\right]}=0\ ,
\end{equation}
and the metric perturbation becomes
\begin{equation}
h_{\mu\nu}=2\eta_{\mu a}E^{a}_{\phantom{a}\nu}\, .
\end{equation}
Now we have all the ingredients to develop the analysis for $f(T,B)$.
%Indeed, the first order perturbation tetrad $E^{a}_{\phantom{a}\mu}$ not is symmetric because the $f\left(T,B\right)$ gravity is not invariant under a local Lorentz trasformation.

%\begin{equation}
%\eta_{(\mu a}E^{a}_{\phantom{a}\nu)}=0
%\end{equation} 

%However, the piece antisymmetric of $E^{a}_{\phantom{a}\mu}$  does not have physical importance because 
%In generale la tetrade e quindi la sua perturbazione al prim'ordine non sono simmetriche essendo la teoria non invariante per trasformazioni locali di Lorentz ovvero
%\begin{equation}
%\eta_{\mu a}E^{a}_{\phantom{a}\nu}\neq \eta_{\nu a}E^{a}_{\phantom{a}\mu}
%\end{equation}
%Tuttavia l'antisimmetria di $E^{a}_{\phantom{a}\mu}$ non ha signficato fisico perch\`e non interviene n\`e nella lagrangiana $L$ che dipende da $T$ e $B$ dove sono presenti solo la parte simmetrica di n\`e nelle equazioni di campo. Per cui possiamo supporre la tetrade simmetrica ponendo uguale a zero la sua parte antisimmetrica ovvero
%\begin{equation}
%\eta_{(\mu a}E^{a}_{\phantom{a}\nu)}=0
%\end{equation} 
%Per la perturbazione metrica si ha:
%\begin{equation}
%h_{\mu\nu}=2\eta_{\mu a}E^{a}_{\phantom{a}\nu}
%\end{equation}
\section{The Weak Field Limit of $f(T,B)$ Teleparallel Gravity }\label{WFLFTB}
Before developing the weak field limit in the context of $f(T,B)$ gravity, let us prove that it  is the teleparallel equivalent of  $f(R)$ gravity \cite{BC}. This statement can be supported by the fact that both theories describe the same physics. It is worth stressing that teleparallel theories are governed by the dynamical variables $e_{a}^{\phantom{a}\mu}$,  components of the tetrad basis $\left\{e_{a}\right\}$. After fixing the  tetrad,  we can express uniquely both the scalar torsion $T$ and the boundary term $B$. Then we can write the scalar torsion $T$ as
\begin{equation}
T=\frac{1}{4}T^{\rho}_{\phantom{\rho}\mu\nu}T_{\rho}^{\phantom{\rho}\mu\nu}+\frac{1}{2}T^{\rho}_{\phantom{\rho}\mu\nu}T^{\nu\mu}_{\phantom{\mu\nu}\rho}-T^{\rho}_{\phantom{\rho}\mu\rho}T^{\nu\mu}_{\phantom{\nu\mu}\nu}\ .
\end{equation} 
From the Weitzenb$\ddot{o}$ck connection $\tilde{\Gamma}^{\rho}_{\ \mu\nu}$,  defined in \eqref{Weitzconn}, and the torsion tensor $T^{\rho}_{\phantom{\rho}\mu\nu}$, defined in \eqref{torsiontensor}, both can be expressed in terms of the tetrad $e_{a}^{\phantom{a}\mu}$.  We get the following expression for the scalar torsion $T$
\begin{equation}
T=e_{a}^{\phantom{a}\mu}\stackrel{\circ}{\nabla^{\nu}}\left(e^{a}_{\phantom{a}\alpha}e^{\phantom{b}\rho}_{b}e^{\phantom{c}\alpha}_{c}\eta^{bc}\right)e^{b}_{\phantom{b}\nu}e^{c}_{\phantom{c}\rho}\eta_{bc}e_{a}^{\phantom{a}\nu}\stackrel{\circ}{\nabla}_{\mu}e^{a}_{\phantom{a}\nu}-e_{a}^{\phantom{a}\mu}\stackrel{\circ}{\nabla}_{\mu}e^{a}_{\phantom{a}\rho}e_{d}^{\phantom{d}\sigma}\stackrel{\circ}{\nabla}_{\sigma}e^{d}_{\phantom{d}\beta}e^{\phantom{b}\beta}_{b}e^{\phantom{c}\rho}_{c}\eta^{bc}\ ,
\end{equation}
where $\stackrel{\circ}{\nabla}_{\mu}$ is a covariant derivative for the  Levi Civita connection $\stackrel{\circ}\Gamma$ given in terms of the tetrad basis. The boundary component $B$, expressed in terms of vierbein, is given  by the relations 
%Le teorie $f(T,B)$ sono equivalenti alle teorie $f(R)$ cio\`e descrivono esattamente la stessa fisica perch\`e sono governate dalle stesse equazioni e non sono teorie pi\`u generali \cite{BC}. Infatti le teorie teleleparallele sono geovernate dalle variabili dinamiche $e_{a}^{\mu}$, le tetradi. Fissata la tetrade cioe il riferimento si possono ricavare sia $T$ che $B$. Infatti dalla definizione della torsione $T$
%\begin{equation}
%T=\left(K^{\mu\nu\rho}K_{\rho\nu\mu}-K^{\rho}K_{\rho}\right)
%\end{equation}
%e dalla connessione di Weitzenb$\ddot{o}$ck $\tilde{\Gamma}^{\rho}_{\ \mu\nu}$, il tensore di torsione $T^{\nu}_{\phantom{\nu}\rho\mu}$, e la contorsione $K^{\nu}_{\phantom{\nu} \rho\mu}$
%\begin{equation}
%5\end{equation}
%\begin{equation}
%T^{\nu}_{\phantom{\nu}\rho\mu}=\tilde{\Gamma}^{\nu}_{\phantom{\nu}\mu\rho}-	\tilde{\Gamma}^{\nu}_{\phantom{\nu}\rho\mu}\ .
%\end{equation}
%\begin{equation}
%K^{\nu}_{\phantom{\nu} \rho\mu}=\tilde{\Gamma}^{\nu}_{\phantom{\nu} \rho\mu}-\stackrel{\circ}{\Gamma}{\hspace{-0.04in}}^{\nu}_{\phantom{\nu} \rho\mu}=\frac{1}{2}\left(T_{\rho\phantom{\nu} \mu}^{\phantom{\rho}\nu}+T_{\mu\phantom{\mu}\rho}^{\phantom{\mu}\nu}-T^{\nu}_{\phantom{\nu}\rho\mu}\right)=e_{a}^{\nu}\stackrel{\circ}{\nabla}_{\mu}e^{a}_{\rho}\ ,
%\end{equation}
%\begin{equation}
%\boxed{
%%%Dalla definizione del {\bf{termine ai bordi}} $B$ in termini della tetrade si ha  
\begin{equation}
B=\frac{2}{e}\partial_{\sigma}\left(eT^{\nu\sigma}_{\phantom{\nu\sigma}\nu}\right)\ ,\qquad T^{\sigma}=e_{\phantom{a}\nu}^{a}\partial^{\sigma}e^{\phantom{a}\nu}_{a}-e^{\phantom{b}\sigma}_{b}e^{\phantom{c}\tau}_{c}\eta^{bc}e^{a}_{\phantom{a}\tau}\partial_{\nu}e^{\phantom{a}\nu}_{a}\ ,
\end{equation}
and then
\begin{equation}
B=\frac{2}{e}\partial_{\sigma}\left[e\left(e_{\phantom{a}\nu}^{a}\partial^{\sigma}e^{\phantom{a}\nu}_{a}-e^{\phantom{b}\sigma}_{b}e^{\phantom{c}\tau}_{c}\eta^{bc}e^{a}_{\phantom{a}\tau}\partial_{\nu}e^{\phantom{a}\nu}_{a}\right)\right]\ .
\end{equation}
Now if we calculate 
\begin{multline}\label{formula3}
-T-B=-\left[e_{a}^{\phantom{a}\mu}\stackrel{\circ}{\nabla^{\nu}}\left(e^{a}_{\phantom{a}\alpha}e^{\phantom{b}\rho}_{b}e^{\phantom{c}\alpha}_{c}\eta^{bc}\right)e^{b}_{\phantom{b}\nu}e^{c}_{\phantom{c}\rho}\eta_{bc}e_{a}^{\phantom{a}\nu}\stackrel{\circ}{\nabla}_{\mu}e^{a}_{\phantom{a}\nu}-e_{a}^{\phantom{a}\mu}\stackrel{\circ}{\nabla}_{\mu}e^{a}_{\phantom{a}\rho}e_{d}^{\phantom{d}\sigma}\stackrel{\circ}{\nabla}_{\sigma}e^{d}_{\phantom{d}\beta}e^{\phantom{b}\beta}_{b}e^{\phantom{c}\rho}_{c}\eta^{bc}\right]\\
-\frac{2}{e}\partial_{\sigma}\left[e\left(e_{\phantom{a}\nu}^{a}\partial^{\sigma}e^{\phantom{a}\nu}_{a}-e^{\phantom{b}\sigma}_{b}e^{\phantom{c}\tau}_{c}\eta^{bc}e^{a}_{\phantom{a}\tau}\partial_{\nu}e^{\phantom{a}\nu}_{a}\right)\right]
\end{multline}
we get exactly the curvature $R$ of the Levi Civita connection $\stackrel{\circ}\Gamma$ expressed in terms of the tetrad basis 
\begin{equation}
R[\stackrel{\circ}{\Gamma}]=e_{a}^{\phantom{a}\theta}e_{b}^{\phantom{b}\nu}\eta^{ab}\left[\partial_{\rho}\stackrel{\circ}\Gamma^{\rho}_{\phantom{\rho}\theta\nu}+\stackrel{\circ}\Gamma^{\rho}_{\phantom{\rho}\sigma\rho}\stackrel{\circ}\Gamma^{\sigma}_{\phantom{\sigma}\theta\nu}-\partial_{\nu}\stackrel{\circ}\Gamma^{\rho}_{\phantom{\rho}\theta\rho}-\stackrel{\circ}\Gamma^{\rho}_{\phantom{\rho}\sigma\nu}\stackrel{\circ}\Gamma^{\sigma}_{\phantom{\sigma}\theta\rho}\right]\ .
\end{equation}
This means that, if we fix the tetrad basis $e_{a}^{\phantom{a}\rho}$ both $T$ and $B$, and therefore  $R$, are uniquely determined. We have no  possibility to disentangle the 3 objects $T$, $B$, and $R$ once the tetrad is given, then  the relation   
%se sostituiamo la connessione di Levi Civita espressa in termini delle tetrade 
%%\stackrel{\circ}{\Gamma}{\hspace{-0.04in}}^{\nu}_{\phantom{\nu} \rho\mu}=\tilde{\Gamma}^{\nu}_{\phantom{\nu} \rho\mu}-\frac{1}{2}\left(T_{\rho\phantom{\nu} \mu}^{\phantom{\rho}\nu}+T_{\mu\phantom{\mu}\rho}^{\phantom{\mu}\nu}-T^{\nu}_{\phantom{\nu}\rho\mu}\right)
%\end{equation}
%otteniamo {\bf{esattamente}} \eqref{formula3}.
%Vuol dire che fissata la base $e_{a}^{\rho}$ non abbiamo alternative si ha sempre
\begin{equation}
R[\stackrel{\circ}\Gamma]=-T-B\,,
\end{equation}
is fixed by the tetrad.
 If $T$ and $B$ were independent,  we could have
\begin{equation}
R[\stackrel{\circ}\Gamma]\neq-T-B\,.
\end{equation}
This would be possible, if we  could express $T$ in a tetrad basis
$e_{(1)a}^{\phantom{(1)a}\rho}$ and $B$ in another tetrad basis $e_{(2)a}^{\phantom{(2)a}\rho}$, %ma questo sarebbe possibile solo se noi esprimessimo $T$ in una base di tetradi  $e_{(1)a}^{\rho}$ e $B$ in un'altra base di tetradi $e_{(2)a}^{\rho}$. 
but this is not possible because both $T$ and $B$ must be expressed in terms of the same basis  $e_{a}^{\phantom{a}\rho}$.  

%Ma cio non \`e possibile perch\`e sia $T$ che $B$ debbono essere espresse in termini delle {\bf{stessa}} base di tetrade $e_{a}^{\rho}$. 
Let us now take into account  the action of $f\left(T, B\right)$ gravity in presence of standard  matter \cite{BC}
\begin{equation}\label{ACTFTB}
S_{f\left(T,B\right)}=\int_{\Omega}d^{4}x\left[\frac{1}{2\kappa^{2}}f\left(T,B\right)+\mathcal{L}_{m}\right]e\ .
\end{equation}
According to the previous considerations, it is that is the teleparallel action equivalent to the  $f(R)$ gravity action.  
The variation of the action \eqref{ACTFTB} with respect to the vierbein fields $e^{a}_{\phantom{a}\rho}$ yields the following field equations 
\begin{multline}\label{EqFielTB}
\frac{4}{e}\partial_{\sigma}\left(ef_{T}S_{a}^{\phantom{a}\rho\sigma}\right)+f\left(T,B\right)e_{a}^{\phantom{a}\rho}-4f_{T}T^{\mu}_{\phantom{\mu}\nu a}S_{\mu}^{\phantom{\mu}\nu\rho}-Bf_{B}e_{a}^{\phantom{a}\rho}
+4\left(\partial_{\lambda}f_{B}\right)S_{a}^{\phantom{a}\lambda\rho}+2e_{a}^{\phantom{a}\rho}\Box f_{B}-2e_{a}^{\phantom{a}\sigma}\nabla_{\sigma}\nabla^{\rho}f_{B}=2\kappa^{2}\mathcal{T}_{a}^{\phantom{a}\rho}\ ,
\end{multline}
where $\mathcal{T}_{a}^{\phantom{a}\rho}$ is the energy momentum tensor of matter defined as 
\begin{equation}
\mathcal{T}_{a}^{\phantom{a}\rho}=-\frac{1}{e}\frac{\delta\left(e\mathcal{L}_{m}\right)}{\delta e^{a}_{\phantom{a}\rho}}\ .
\end{equation}
Supposing $f\left(T,B\right)$ being an analytic function of $T$ and $B$ we can expand it as
\begin{multline}
f\left(T,B\right)=f\left(0\right)+f_{T}\left(0\right)T+f_{B}\left(0\right)B+f_{TB}\left(0\right)TB
+f_{TT}\left(0\right)T^{2}+f_{BB}\left(0\right)B^{2}+\mathcal{O}\left(TB, T^2, B^2\right)\ .
\end{multline}
The linearized field equations are 
\begin{equation}\label{EqFielTBLin}
4f_{T}\left(0\right)\partial_{\sigma}S_{\tau}^{\phantom{\tau}\rho\sigma\left(1\right)}+4f_{B^2}\left(0\right)\delta^{\rho}_{\tau}\Box B^{\left(1\right)}-4f_{B^2}\left(0\right)\partial_{\tau}\partial^{\rho}B^{\left(1\right)}=2\kappa^{2}\mathcal{T}_{\tau}^{\phantom{\tau}\rho\left(0\right)}\ .
\end{equation}
The field equations \eqref{EqFielTBLin} are gauge-invariant, namely  under transformations of gauge 
\begin{equation}
E_{\mu\nu}\longrightarrow E_{\mu\nu}+\partial_{\mu}\Lambda_{\nu}+\partial_{\nu}\Lambda_{\mu}\ ,
\end{equation} 
remain invariant to first order, with $\Lambda_{\mu}$ infinitesimal. We can use  the   Lorentz gauge
\begin{equation}
\partial_{\mu}\left(E^{\mu\nu}-\frac{1}{2}\eta^{\mu\nu}E\right)=0\ ,
\end{equation}
where we set
\begin{equation}
E_{\mu\nu}=\eta_{\mu a}E^{a}_{\phantom{a}\nu}\ , \qquad E=\delta_{a}^{\phantom{a}\mu}E^{a}_{\phantom{a}\mu}\ .
\end{equation}
In the harmonic gauge,  $B^{\left(1\right)}$ and  Eq.\eqref{DerSupPot} take the form
\begin{equation}
B^{\left(1\right)}=-\Box\bar{E}\ ,
\end{equation}
\begin{equation}
2\partial_{\nu}S_{\rho}^{\phantom{\rho}\mu\nu(1)}=-\Box\bar{E}^{\mu}_{\phantom{\rho}\rho}\ .
\end{equation}
Thus the Eq.\eqref{EqFielTBLin} becomes in a simpler form
\begin{equation}\label{WFLEFFTB}
\boxed{
f_{T}^{(0)}\Box \bar{E}^{\rho}_{\phantom{\rho}\tau}+2f_{B^2}^{(0)}\left(\delta^{\rho}_{\tau}\Box^{2}-\partial_{\tau}\partial^{\rho}\Box\right)\bar{E}=-2\kappa^{2}\mathcal{T}^{\phantom{\tau}\rho\left(0\right)}_{\tau}
}\ ,
\end{equation}
where we called $\bar{E}_{\mu\nu}$
\begin{equation}\label{newfieldgaugelorentz}
\bar{E}_{\mu\nu}=E_{\mu\nu}-\frac{1}{2}\eta_{\mu\nu}E\ .
\end{equation}
Hence  the trace of Eq. \eqref{WFLEFFTB} is
\begin{equation}\label{EqtraceFTBlin}
\boxed{
f_{T}^{\left(0\right)}\Box\bar{E}+6f_{B^2}^{\left(0\right)}\Box^{2}\bar{E}=-2\kappa^{2}\mathcal{T}^{\left(0\right)}
}\ .
\end{equation}
We have now all the ingredients to develop the GW theory for $f(T,B)$ teleparallel gravity.

\section{Gravitational Waves in  $f(T,B)$ Teleparallel Gravity}\label{GWTP}

Let us now derive the GWs for  $f(T,B)$ gravity in vacuum as  solutions of Eq.\eqref{WFLEFFTB}. We start from the trace equation \eqref{EqtraceFTBlin} in vacuum 
\begin{equation}\label{EqtraceFTBlinVacuum}
f_{T}^{\left(0\right)}\Box\bar{E}+6f_{B^2}^{\left(0\right)}\Box^{2}\bar{E}=0\ ,
\end{equation}
that, in the $k$-space, becomes the algebraic equation \cite{CCC} {\bf with $f_{B^2}^{\left(0\right)}\neq 0$}
\begin{equation}\label{eqtraceTBkSpace}
\left(k^{4}-\frac{f_{T}^{\left(0\right)}}{6 f_{B^2}^{\left(0\right)}}k^2\right)\hat A\left(k_{0},\mathbf{k}\right)=0\ ,
\end{equation}
where $k^{2}=\omega^{2}-\mathbf{k}\cdot\mathbf{k}=\omega^{2}-q^{2}$. Here  $k^{\mu}=\left(\omega,\bf{k}\right)$ is the wave four-vector.  If $f_{B^2}^{\left(0\right)}§=0$, Eq. \eqref{EqFielTBLin} becomes
\begin{equation}
2f_{T}\left(0\right)\partial_{\sigma}S_{\tau}^{\phantom{\tau}\rho\sigma\left(1\right)}=\kappa^{2}\mathcal{T}_{\tau}^{\phantom{\tau}\rho\left(0\right)}\ ,
\end{equation}
that is the linearized field equations of $f\left(T\right)$ gravity with matter. In the harmonic gauge, we obtain 
\begin{equation}\label{WFLEFFTBFB2=0}
f_{T}^{(0)}\Box \bar{E}^{\rho}_{\phantom{\rho}\tau}=-2\kappa^{2}\mathcal{T}^{\phantom{\tau}\rho\left(0\right)}_{\tau}
\ ,
\end{equation}
whose  trace equation is
\begin{equation}\label{EqtraceFTBlinFB2=0}
f_{T}^{\left(0\right)}\Box\bar{E}=-2\kappa^{2}\mathcal{T}^{\left(0\right)}\ .
\end{equation}
The solutions of  Eq. \eqref{EqtraceFTBlinFB2=0} are the gravitational waves of  $f(T)$ gravity whose polarizations are the two standard $+$  and $\times$ modes of GR, as demonstrated in \cite{bamba}.
Therefore the $f(T,B)$ gravity, for $f_{B^2}^{\left(0\right)}§=0$, reproduces the results of $f(T)$ gravity. 

The general solution of Eq.\eqref{EqtraceFTBlinVacuum} can be  expressed as a  Fourier integral 
\begin{equation}
\bar{E}\left(t,\mathbf{x}\right)=\int \frac{d^{4}k}{\left(2\pi\right)^{2}}\hat A\left(k_{0}, \mathbf{k}\right)e^{i k^{\alpha}x_{\alpha}}\ .
\end{equation}
Then, we obtain two solutions of Eq.\eqref{eqtraceTBkSpace} for $A\left(k_{0}, \mathbf{k}\right)\neq 0$, that is  
\begin{equation}
k_{1}^{2}=0\ , \qquad \text{and} \qquad k_{2}^{2}=\frac{f_{T}^{\left(0\right)}}{6 f_{B^2}^{\left(0\right)}}\neq 0\ ,
\end{equation} and the  integral of  trace equation \eqref{EqtraceFTBlinVacuum} in vacuum is
\begin{equation}\label{soltracevacuum}
\bar{E}\left(t,\mathbf{x}\right)=\sum_{m=1}^{2}\int \frac{d^{3}\mathbf{k}}{\left(2\pi\right)^{3/2}}\left(\hat{A}_{m}\left(\mathbf{k}\right)e^{i k_{m}^{\alpha}x_{\alpha}}+c. c.\right) \ .
\end{equation}
Therefore substituting Eq.\eqref{soltracevacuum} into Eq.\eqref{WFLEFFTB}, in vacuum, we get 
\begin{equation}
\Box\bar{E}_{\rho\tau}\left(x\right)=\int \frac{d^{3}\mathbf{k}}{(2\pi)^{3/2}}\left\{\left(-\frac{k_{2}^{2}}{3}\right)\left[\eta_{\rho\tau}-\frac{\left(k_{2}\right)_{\rho}\left(k_{2}\right)_{\tau}}{k_{2}^{2}}\right]\right\}\left(\hat{A}_{2}\left(\mathbf{k}\right)e^{ik_{2}^{\alpha}x_{\alpha}}+c.c.\right)\ .
\end{equation}
Finally, the general solution of Eq.\eqref{WFLEFFTB}, in vacuum, expressed as a homogeneous plus a particular solution is
\begin{align}\label{GWEFTBTG1}
\bar{E}_{\rho\tau}\left(x\right)&=\int \frac{d^{3}\mathbf{k}}{(2\pi)^{3/2}}\hat{C}_{\rho\tau}\left(\mathbf{k}\right)e^{ik_{1}^{\alpha}x_{\alpha}} \nonumber\\
&+\int \frac{d^{3}\mathbf{k}}{(2\pi)^{3/2}}\left\{\left(-\frac{1}{3}\right)\left[\eta_{\rho\tau}-\frac{\left(k_{2}\right)_{\rho}\left(k_{2}\right)_{\tau}}{k_{2}^{2}}\right]\right\}\hat{A}_{2}\left(\mathbf{k}\right)e^{ik_{2}^{\alpha}x_{\alpha}}+c.c.\ .
\end{align}
From Eq.\eqref{newfieldgaugelorentz}, it is possible to derive the GWs for $f\left(T, B\right)$ gravity, that is
\begin{align}\label{GWEFTBTG}
E_{\rho\tau}\left(x\right)&=\int \frac{d^{3}\mathbf{k}}{(2\pi)^{3/2}}\hat{C}_{\rho\tau}\left(\mathbf{k}\right)e^{ik_{1}^{\alpha}x_{\alpha}} \nonumber\\
&+\int \frac{d^{3}\mathbf{k}}{(2\pi)^{3/2}}\left\{\frac{1}{3}\left[\frac{\eta_{\rho\tau}}{2}+\frac{\left(k_{2}\right)_{\rho}\left(k_{2}\right)_{\tau}}{k_{2}^{2}}\right]\right\}\hat{A}_{2}\left(\mathbf{k}\right)e^{ik_{2}^{\alpha}x_{\alpha}}+c.c.\ .
\end{align}
Starting from this solution we can analyze the polarizations and the helicity of GWs.

\section{Polarizations and helicity}\label{POLHEL}
A useful way to visualize the polarizations of  gravitational waves is to derive the geodesic deviation that they generate via the equation for geodesic deviation. Let us consider the wave propagating in  $+\hat{z}$ direction, in a local proper reference frame, and take into account the equation for geodesic deviation 
\begin{equation}\label{eqdevgeoelectric}
\ddot x^{i}=-R^{i}_{\phantom{i}0k0}x^{k}\ ,
\end{equation}
where the Latin index range over the set $\left\{1,2,3\right\}$ and $R^{i}_{\phantom{i}0k0}$ are so-called "electric" components of the Riemann tensor, the only measurable components \cite{STRAGR}.
Substituting the linearized electric components of the Riemann tensor $R^{\left(1\right)}_{\phantom{1}i0j0}$,  expressed in terms of the tetrad perturbation $E_{\mu\nu}$,  
\begin{equation}
R^{\left(1\right)}_{\phantom{1}i0j0}=\left(E_{i0,j0}+E_{j0,i0}-E_{ij,00}-E_{00,ij}\right)\ ,
\end{equation}
into Eq.\eqref{eqdevgeoelectric}, we obtain
\begin{equation}\label{eqdevgeolinear}
\begin{cases}
\ddot x(t)=-\left(xE_{11,00}+yE_{12,00}\right) \\
\ddot y(t)=- \left(xE_{12,00}+yE_{11,00}\right)\\
\ddot z(t)=\left(2E_{03,03}-E_{33,00}-E_{00,33}\right)z
\end{cases}\ .
\end{equation}
From Eq.\eqref{GWEFTBTG}, for $k_{1}^{2}=0$ the massless plane wave, travelling in  $+\hat{z}$ direction, whose propagation speed is equal to $c$, keeping $\mathbf{k}$ fixed and $k_{1}^{\mu}=\left(\omega_{1},0,0,k_{z}\right)$, we have   
\begin{equation}\label{FirstOrderTetradmassless}
E^{(k_{1})}_{\mu\nu}\left(t,z\right)=\sqrt{2}\left[\hat{\epsilon}^{(+)}\left(\omega_{1}\right)\epsilon^{(+)}_{\mu\nu}+\hat{\epsilon}^{(\times)}\left(\omega_{1}\right)\epsilon^{(\times)}_{\mu\nu}\right]e^{i\omega_{1}\left(t-z\right)}+c.c.\ ,
\end{equation}
where 
\begin{equation}
\epsilon^{(+)}_{\mu\nu}=\frac{1}{\sqrt{2}}
\begin{pmatrix} 
0 & 0 & 0 & 0 \\
0 & 1 & 0 & 0 \\
0 & 0 & -1 & 0 \\
0 & 0 & 0 & 0
\end{pmatrix}\ .
\end{equation}
\begin{equation}
\epsilon^{(\times)}_{\mu\nu}=\frac{1}{\sqrt{2}}
\begin{pmatrix} 
0 & 0 & 0 & 0 \\
0 & 0 & 1 & 0 \\
0 & 1 & 0 & 0 \\
0 & 0 & 0 & 0
\end{pmatrix}\ ,
\end{equation}
and $\omega_{1}=k_{z}$. Furthermore, always from Eq.\eqref{GWEFTBTG} for $k_{2}^{2}\neq 0$, the massive plane wave propagating in  $+\hat{z}$ direction, is 
\begin{multline}\label{FirstOrderTetradmassive}
E^{(k_{2})}_{\mu\nu}\left(t,z\right)=\Biggl[\left(\frac{1}{2}+\frac{\omega^{2}_{2}}{k_{2}^{2}}\right)\epsilon_{\mu\nu}^{(TT)}-\frac{\sqrt{2}\omega_{2}k_{z}}{k_{2}^{2}}\epsilon_{\mu\nu}^{(TS)}
-\frac{1}{\sqrt{2}}\epsilon_{\mu\nu}^{(b)}+\left(\frac{k_{z}^{2}}{k_{2}^{2}}-\frac{1}{2}\right)\epsilon_{\mu\nu}^{(l)}\Biggr]\frac{\hat{A}_{2}\left(k_{z}\right)}{3}e^{i\left(\omega_{2}t-k_{z}z\right)}+c.c.\ .
\end{multline}
Here, the propagation speed is less then $c$, keeping $\mathbf{k}$ fixed and $k_{2}^{\mu}=\left(\omega_{2},0,0,k_{z}\right)$. The polarizations are   
\begin{align}
\epsilon^{(TT)}_{\mu\nu}&=
\begin{pmatrix} 
1 & 0 & 0 & 0 \\
0 & 0 & 0 & 0 \\
0 & 0 & 0 & 0 \\
0 & 0 & 0 & 0
\end{pmatrix}\ , &
\epsilon^{(TS)}_{\mu\nu}&=\frac{1}{\sqrt{2}}
\begin{pmatrix} 
0 & 0 & 0 & 1 \\
0 & 0 & 0 & 0 \\
0 & 0 & 0 & 0 \\
1 & 0 & 0 & 0
\end{pmatrix}\ ,\\
\epsilon^{(b)}_{\mu\nu}&=\frac{1}{\sqrt{2}}
\begin{pmatrix} 
0 & 0 & 0 & 0 \\
0 & 1 & 0 & 0 \\
0 & 0 & 1 & 0 \\
0 & 0 & 0 & 0
\end{pmatrix}\ , &
\epsilon^{(l)}_{\mu\nu}&=
\begin{pmatrix} 
0 & 0 & 0 & 0 \\
0 & 0 & 0 & 0 \\
0 & 0 & 0 & 0 \\
0 & 0 & 0 & 1
\end{pmatrix}\ .
\end{align}
In more compact form, the tetrad linear perturbation  $E_{\mu\nu}$,  traveling in the $+\hat{z}$ direction assuming  $\mathbf{k}$ fixed,  is
\begin{equation} 
E_{\mu\nu}\left(t,z\right)=\sqrt{2}\left[\hat{\epsilon}^{(+)}\left(\omega_{1}\right)\epsilon^{(+)}_{\mu\nu}+\hat{\epsilon}^{(\times)}\left(\omega_{1}\right)\epsilon^{(\times)}_{\mu\nu}\right]e^{i\omega_{1}\left(t-z\right)}+\hat{\epsilon}^{\left(s\right)}_{\mu\nu}\left(k_{z}\right)e^{i\left(\omega_{2}t-k_{z}z\right)}+c.c.\ ,
\end{equation}
where $\hat{\epsilon}^{\left(s\right)}_{\mu\nu}$ is the polarization tensor associated to the scalar mode 
\begin{equation}
\hat{\epsilon}^{\left(s\right)}_{\mu\nu}\left(k_{z}\right)=\Biggl[\left(\frac{1}{2}+\frac{\omega^{2}_{2}}{k_{2}^{2}}\right)\epsilon_{\mu\nu}^{(TT)}-\frac{\sqrt{2}\omega_{2}k_{z}}{k_{2}^{2}}\epsilon_{\mu\nu}^{(TS)}\\
-\frac{1}{\sqrt{2}}\epsilon_{\mu\nu}^{(b)}+\left(\frac{k_{z}^{2}}{k_{2}^{2}}-\frac{1}{2}\right)\epsilon_{\mu\nu}^{(l)}\Biggr]\frac{\hat{A}_{2}\left(k_{z}\right)}{3}\ .
\end{equation}
The only degree of freedom $\hat{A}_{2}$ produces the scalar polarization $\hat{\epsilon}^{\left(s\right)}_{\mu\nu}$. It is worth noticing that, for  $f(R)$ gravity, we have the  three d.o.f.: $\hat{\epsilon}^{(+)}$, $\hat{\epsilon}^{(\times)}$ and $\hat{A}_{2}$ \cite{YGSH, LGHL, IKEDA}. The scalar mode is a combination of longitudinal and breathing scalar modes. In fact,  the polarization tensor $\hat{\epsilon}^{\left(s\right)}_{\mu\nu}$, restricted to  spatial components $\hat{\epsilon}^{\left(s\right)}_{i,j}$, is provided by 
\begin{equation}
\hat{\epsilon}^{\left(s\right)}_{i,j}=-\frac{1}{3\sqrt{2}}\hat{A}_{2}\left(k_{z}\right)\epsilon_{i,j}^{(b)}+\frac{1}{3}\left(\frac{k_{z}^{2}}{k_{2}^{2}}-\frac{1}{2}\right)\hat{A}_{2}\left(k_{z}\right)\epsilon_{i,j}^{(l)}\ ,
\end{equation}
where $(i,j)$ range over $(1,2,3)$. Hence, for massless plane wave $E_{\mu\nu}^{(k_{1})}$, Eqs.\eqref{eqdevgeolinear} give
\begin{equation}
\begin{cases}

\ddot x(t)=\omega_{1}^{2}\left[\hat{\epsilon}^{\left(+\right)}\left(\omega_{1}\right)x+\hat{\epsilon}^{\left(\times\right)}\left(\omega_{1}\right)y\right]e^{i\omega_{1}\left(t-z\right)}+c.c. \\ \\
\ddot y(t)=\omega_{1}^{2}\left[\hat{\epsilon}^{\left(\times\right)}\left(\omega_{1}\right)x-\hat{\epsilon}^{\left(+\right)}\left(\omega_{1}\right)y\right]e^{i\omega_{1}\left(t-z\right)}+c.c.\\ \\
\ddot z(t)=0
\end{cases}\ ,
\end{equation}
where we obtain the two standard 
polarizations of GR, the purely transverse plus and the cross polarization, two-helicity massless tensor modes. \\
Instead, for a massive plane wave $E^{(k_{2})}$ with $M^2=\omega^{2}_{2}-k^{2}_{z}$, the  geodesic deviation Eq.\eqref{eqdevgeolinear} becomes
\begin{equation}
\begin{cases}
\ddot x(t)=-\frac{1}{6}\omega^{2}_{2}\hat{A}_{2}\left(k_{z}\right)x e^{i\left(\omega_{2}t-k_{z}z\right)}+c.c.\\ \\
\ddot y(t)=-\frac{1}{6}\omega^{2}_{2}\hat{A}_{2}\left(k_{z}\right)y e^{i\left(\omega_{2}t-k_{z}z\right)}+c.c.\\ \\
\ddot z(t)=-\frac{1}{6}M^{2}\hat{A}_{2}\left(k_{z}\right)ze^{i\left(\omega_{2}t-k_{z}z\right)}+c.c.
\end{cases}\ .
\end{equation}
This system of equations can be integrated  assuming that $E_{\mu\nu}\left(t,z\right)$ is small. Hence we have 
\begin{equation}
\begin{cases}
x(t)=x(0)+\frac{1}{6}\hat{A}_{2}\left(k_{z}\right)x(0)e^{i\left(\omega_{2}t-k_{z}z\right)}+c.c.\\ \\
y(t)=y(0)+\frac{1}{6}\hat{A}_{2}\left(k_{z}\right)y(0)e^{i\left(\omega_{2}t-k_{z}z\right)}+c.c.\\ \\
z(t)=z(0)+\frac{1}{6\omega^{2}_{2}}M^{2}\hat{A}_{2}\left(k_{z}\right)z(0)e^{i\left(\omega_{2}t-k_{z}z\right)}+c.c.
\end{cases}\ .
\end{equation}
When a GW strikes a sphere of particles of radius $r=\sqrt{x^2(0)+y^{2}(0)+z^{2}(0)}$, this  will be distorted into an ellipsoid described by
\begin{equation}
\left(\frac{x}{\rho_{1}(t)}\right)^{2}+\left(\frac{y}{\rho_{1}(t)}\right)^{2}+\left(\frac{z}{\rho_{2}(t)}\right)^{2}=r^{2}\ ,
\end{equation}
where $\rho_{1}(t)=1+\frac{1}{3}\hat{A}_{2}\left(k_{z}\right)\cos\left(\omega_{2}t-k_{z}z\right)$ and $\rho_{2}(t)=1+\frac{M^{2}}{3\omega_{2}^{2}}\hat{A}_{2}\left(k_{z}\right)\cos\left(\omega_{2}t-k_{z}z\right)$ both varying between their maximum and minimum values. This swinging ellipsoid represents an additional scalar polarization, zero-helicity which is partly longitudinal and partly
transverse \cite{PW}. 

According to these considerations, the d.o.f. of  $f(T,B)$ gravity are three: two of these, $\hat{\epsilon}^{\left(+\right)}$ and $\hat{\epsilon}^{\left(\times\right)}$, generate the tensor modes while the degree of freedom $\hat{A}_{2}$ generates the mixed scalar mode. In summary,  $f\left(T,B\right)$ gravity has three polarizations namely, two tensor modes and one mixed scalar mode exactly like $f(R)$ gravity (see \cite{IKEDA} for a discussion).

It is possible to  derive the same results  adopting the Newman-Penrose (NP) formalism. It is not directly applicable to  massive waves because it was, in origin,   worked out  for  massless waves. However, it is possible  to  adopt its generalization to massive waves  propagating along non-null geodesics \cite{NP}. It is worth noticing that the  little group $E\left(2\right)$ classification fails for massive waves. One can  introduce a local quasi-normal  null tetrad basis $\left(k, l, m, \bar{m}\right)$ as 
\begin{align}
k&=\frac{1}{\sqrt{2}}\left(\partial_{t}+\partial_{z}\right)\ ,
& l&=\frac{1}{\sqrt{2}}\left(\partial_{t}-\partial_{z}\right)\ ,\\
m&=\frac{1}{\sqrt{2}}\left(\partial_{x}+i\partial_{y}\right)\ ,
& \bar{m}&=\frac{1}{\sqrt{2}}\left(\partial_{x}-i\partial_{y}\right)\ ,
\end{align}
which satisfies the relations 
\begin{gather}
k\cdot l=-m\cdot\bar{m}=1\ , \nonumber\\
k\cdot k=l\cdot l=m\cdot m=\bar{m}\cdot\bar{m}=0\ ,\\
k\cdot m=k\cdot\bar{m}=l\cdot m=l\cdot\bar{m}=0\ . \nonumber
\end{gather}
Let us consider the four-dimensional Weyl tensor $C_{\mu\nu\rho\sigma}$ defined as
\begin{equation}
C_{\mu\nu\rho\sigma}=R_{\mu\nu\rho\sigma}-2g_{[\mu|[\rho}R_{\sigma]|\nu]}+\frac{1}{3}g_{\mu[\rho}g_{\sigma]\nu}R\ .
\end{equation}
The five complex Weyl-NP scalars, classified by the  spin weight $s$, can be expressed from the Weyl tensor   in a null tetrad basis as
%\begin{equation}
\begin{eqnarray}
s=+2  & \Psi_{0}\equiv C_{kmkm}\ ,\nonumber\\
s=+1  & \Psi_{1}\equiv C_{klkm}\ ,\nonumber\\
s=0  & \Psi_{2}\equiv C_{km\bar{m}}\ ,\\
s=-1  & \Psi_{3}\equiv C_{kl\bar{m}l}\ ,\nonumber\\
s=-2  & \Psi_{4}\equiv C_{\bar{m}l\bar{m}l}\ ,\nonumber
\end{eqnarray}
while the ten Ricci-NP scalars, can be expressed from Ricci tensor  in a null tetrad basis as
\begin{eqnarray*}
s=2 & \Phi_{02}\equiv-\frac{1}{2}R_{mm\ ,}\\
\end{eqnarray*} 
\begin{equation*}
s=1\quad 
\begin{cases}
\Phi_{01}\equiv-\frac{1}{2}R_{km}\\
\Phi_{12}\equiv-\frac{1}{2}R_{lm}\\
\end{cases}\ ,
\end{equation*}
\begin{equation*}
s=0\quad
\begin{cases}
\Phi_{00}\equiv-\frac{1}{2}R_{kk}\\
\Phi_{11}\equiv-\frac{1}{4}\left(R_{kl}+R_{m\bar{m}}\right)\\
\Phi_{22}\equiv-R_{ll}
\end{cases}\ ,
\end{equation*} 
\begin{equation*}
s=-1\quad
\begin{cases}
\Phi_{10}\equiv-\frac{1}{2}R_{k\bar{m}}=\Phi_{01}^{*}\\
\Phi_{21}\equiv-\frac{1}{2}R_{l\bar{m}}=\Phi_{21}^{*}\\
\end{cases}\ ,
\end{equation*} 
\begin{eqnarray}
s=-2 & \Phi_{20}\equiv-\frac{1}{2}R_{m\bar{m}}=\Phi_{02}^{*}\ ,\nonumber\\
& \Lambda=\frac{R}{24}\ .
\end{eqnarray} 
The driving- force matrix $S(t)$ can be expressed in terms of the six new basis polarization matrices $W_{A}(\mathbf{e}_{z})$ along the  wave direction $\hat{\mathbf{k}}=\mathbf{e}_{z}$, where the index $A$ ranges over $\{1,2,3,4,5,6\}$ \cite{ELL,ELLWW}. It is 
\begin{equation}
S\left(t\right)=\sum_{A}p_{A}\left(\mathbf{e}_{z},t\right)W_{A}\left(\mathbf{e}_{z}\right)\ ,
\end{equation}
where 
\begin{align}
W_{1}\left(\mathbf{e}_{z}\right)=&-6\begin{pmatrix} 
0 & 0 & 0 \\
0 & 0 & 0 \\
0 & 0 & 1
\end{pmatrix}\ , & W_{2}\left(\mathbf{e}_{z}\right)=&-2\begin{pmatrix} 
0 & 0 & 1 \\
0 & 0 & 0 \\
1 & 0 & 0
\end{pmatrix}\ ,\nonumber\\
W_{3}\left(\mathbf{e}_{z}\right)=&2\begin{pmatrix} 
0 & 0 & 0 \\
0 & 0 & 1 \\
0 & 1 & 0
\end{pmatrix}\ , & W_{4}\left(\mathbf{e}_{z}\right)=&-\frac{1}{2}\begin{pmatrix} 
1 & 0 & 0 \\
0 & -1 & 0 \\
0 & 0 & 0
\end{pmatrix}\ ,\nonumber\\
W_{5}\left(\mathbf{e}_{z}\right)=&\frac{1}{2}\begin{pmatrix} 
0 & 1 & 0 \\
1 & 0 & 0 \\
0 & 0 & 0
\end{pmatrix}\ , & W_{6}\left(\mathbf{e}_{z}\right)=&-\frac{1}{2}\begin{pmatrix} 
1 & 0 & 0 \\
0 & 1 & 0 \\
0 & 0 & 0
\end{pmatrix}\ .
\end{align}
Here $p_{A}\left(\mathbf{e}_{z},t\right)$ are the amplitudes of the wave \cite{MN, WILL, AMA, MY, BCLN}. Taking into account that the spatial components of matrix $S(t)$ are the electric components of Riemann tensor, we have
\begin{equation}
S_{ij}(t)=R_{i0j0}\ .
\end{equation}
The amplitudes of six polarizations can be expressed both in terms of the electric components of the Riemann tensor $R_{i0j0}$, and in Weyl and Ricci scalars \cite{HKL, BEMI, AMdeA, dePMM, WAG, FSGS, YGSH}, that is
\begin{align}
p_{1}^{\left(l\right)}=&-\frac{1}{6}R_{0303}=-\frac{1}{3}\left[\mathrm{Re}\left(\Psi_{2}\right)+\Phi_{11}-\Lambda\right]\ ,\nonumber\\
p_{2}^{\left(x\right)}=&-\frac{1}{2}R_{0301}=-\frac{1}{2}\left[-\mathrm{Re}\left(\Psi_{1}\right)+\mathrm{Re}\left(\Psi_{3}\right)-\mathrm{Re}\left(\Phi_{01}\right)+\mathrm{Re}\left(\Phi_{12}\right)\right]\ ,\nonumber\\
p_{3}^{\left(y\right)}=&\frac{1}{2}R_{0302}=\frac{1}{2}\left[-\mathrm{Im}\left(\Psi_{1}\right)-\mathrm{Im}\left(\Psi_{3}\right)-\mathrm{Im}\left(\Phi_{01}\right)+\mathrm{Im}\left(\Phi_{12}\right)\right]\ ,\nonumber\\
p_{4}^{\left(+\right)}=&-R_{0101}+R_{0202}=-\mathrm{Re}\left(\Psi_{0}\right)-\mathrm{Re}\left(\Psi_{4}\right)-2\mathrm{Re}\left(\Psi_{02}\right)\ ,\nonumber\\
p_{5}^{\left(\times\right)}=&2R_{0102}=\mathrm{Im}\left(\Psi_{0}\right)-\mathrm{Im}\left(\Psi_{4}\right)-2\mathrm{Im}\left(\Psi_{02}\right)\ ,\nonumber\\
p_{6}^{\left(b\right)}=&-R_{0101}-R_{0202}=2\mathrm{Re}\left(\Psi_{2}\right)-\Phi_{00}-\Phi_{22}+4\Lambda\ ,
\end{align}
where the six polarizations modes are: the longitudinal mode $p_{1}^{\left(l\right)}$, the vector-$x$ mode $p_{2}^{\left(x\right)}$, the vector-$y$ mode $p_{3}^{\left(y\right)}$, the plus mode $p_{4}^{\left(+\right)}$, the cross mode $p_{5}^{\left(\times\right)}$, and the breathing mode $p_{6}^{\left(b\right)}$.
Under the Lorentz gauge and by Eqs.\eqref{FirstOrderTetradmassless} and \eqref{FirstOrderTetradmassive} for non-null geodesic congruences of gravitational waves, traveling along the $+\hat{z}$ direction,  we obtain the six polarization amplitudes $p_{A}\left(\mathbf{e}_{z}, t\right)$
\begin{align*}
p_{1}^{\left(l\right)}&=-\frac{1}{6}\left[\frac{\omega^{2}-k_{z}^{2}}{\omega^{2}+k_{z}^{2}}\left(\omega^{2}E_{33}-k_{z}^{2}E_{00}\right)\right]\ ,\\
p_{2}^{\left(x\right)}&=-\frac{1}{2}\left(\omega^{2}-k_{z}^{2}\right)E_{13}\ ,\\
p_{3}^{\left(y\right)}&=\frac{1}{2}\left(\omega^{2}-k_{z}^{2}\right)E_{23}\ ,\\
p_{4}^{\left(+\right)}&=\left(\frac{\omega^{2}-k_{z}^{2}}{\omega^{2}+k_{z}^{2}}\right)\omega^{2}\left(E_{00}+E_{33}\right)+2\omega^{2}E_{22}\ ,\\
p_{5}^{\left(\times\right)}&=2\omega^{2}E_{12}\ ,\\
p_{6}^{\left(b\right)}&=\left(\frac{\omega^{2}-k_{z}^{2}}{\omega^{2}+k_{z}^{2}}\right)\omega^{2}\left(E_{00}+E_{33}\right)\ .
\end{align*}
Finally we get for a massless mode $\omega_{1}$ and massive mode $\omega_{2}$, keeping $\mathbf{k}$ fixed,  the following amplitudes
\begin{align}\label{masspolarampl}
p_{1}^{\left(l\right)}\left(t, z\right)&=\frac{1}{36}\left(\omega_{2}^{2}-k_{z}^{2}\right)\hat{A}_{2}\left(k_{z}\right)e^{i\left(\omega_{2}t-k_{z}z\right)}+c.c.\ ,\nonumber\\
p_{2}^{\left(x\right)}\left(t, z\right)&=p_{3}^{\left(y\right)}\left(t, z\right)=0\ ,\nonumber\\
p_{4}^{\left(+\right)}\left(t, z\right)&=-2\omega_{1}^{2}\hat{\epsilon}^{\left(+\right)}\left(\omega_{1}\right)e^{i\omega_{1}\left(t-z\right)}+c.c.\ ,\nonumber\\
p_{5}^{\left(\times\right)}\left(t, z\right)&=2\omega_{1}^{2}\hat{\epsilon}^{\left(\times\right)}\left(\omega_{1}\right)e^{i\omega_{1}\left(t-z\right)}+c.c.\ ,\nonumber\\
p_{6}^{\left(b\right)}\left(t, z\right)&=\frac{\omega_{2}^{2}}{3}\hat{A}_{2}\left(k_{z}\right)e^{i\left(\omega_{2}t-k_{z}z\right)}+c.c.\ .
\end{align}
It is evident, from Eqs.\eqref{masspolarampl},  that the two vector modes $p_{2}^{\left(x\right)}$ and $p_{3}^{\left(y\right)}$ are suppressed while the two standard plus and cross  transverse tensor polarization modes $p_{4}^{\left(+\right)}$ and $p_{5}^{\left(\times\right)}$  survive together with the two longitudinal and transverse breathing scalar modes $p_{1}^{\left(l\right)}$ and $p_{6}^{\left(b\right)}$. However only one degree of freedom $\hat{A}_{2}$ intervenes in both $b$ and $l$ scalar modes, giving rise to their mixed state  $s$. This reduces polarizations from four to three.

\section{Discussion and Conclusions}
\label{conc}
TEGR is equivalent to Einstein's GR because they are two representations of the same dynamics.  This is not true for their extensions $f(T)$ and 
$f(R)$ theories and, in  general, for  higher order gravity theories constructed by the torsion $T$ and curvature $R$ scalars  \cite{bamba}. To restore  the equivalence, we must take into account the boundary term $B$ which relates $T$ and $R$. Because $T$ and $B$ are derived from the same tetrad, $R$ is univocally defined, so $f(T,B)\equiv f(R)$ according to dynamics.

Thus, we have obtained the exact field equations of $f\left(T, B \right)$ gravity (in presence of matter) and then we have linearized them in the low energy limit. This allows to get  gravitational waves and then to study their polarization and  helicity. To this end, one can adopt  the geodesic deviation and the NP formalism. 
 
 In this framework. it is possible to show that both  $f(R)$ gravity and  $f(T,B)$ teleparallel gravity  have three polarizations \cite{YGSH, LGHL, MY, BCLN,CCDLV}. The third polarization, with respect to the standard $\times$ and $+$ of GR, emerges as a combination of longitudinal and  breathing scalar modes. 
 
 Same authors claims that the polarization of $f(R)$ theory are four because count two scalar polarizations instead of the single scalar state \cite{KPJ}. This is true for a gravitational wave without mass, where the longitudinal and transverse breathing modes are independent of each other. However,  for a massive gravitational wave,  both longitudinal  and breathing modes are determined by a unique degree of freedom, i.e. $\hat{A}_{2}$ and  then cannot be, in principle,  disentangled. According to this result, they can  be combined  giving only a scalar mode. 
  
In  $f\left(T, B\right)$  teleparallel gravity, the presence of a massive scalar mode mixes the transverse breathing and the longitudinal modes, in addition to the two standard massless  tensor polarizations. This further term  is due to the boundary terms $B$, which survives to first order in $E^{a}_{\phantom{a}\mu}$. It is worth stressing that in $f(T)$ gravity only the two standard modes of GR are present \cite{bamba}. 
  
More precisely, it is the first order boundary terms $B^{(1)}$ that generates the massive scalar mode and then adds, to the 2-spin massless tensor modes of GR,  an extra 0-spin massive scalar mode. Furthermore,  as it is well known,  $f(R)$ gravity is equivalent to a scalar-tensor theory \cite{OHNLON, TT}. It means that under a conformal transformation,  it is equivalent to GR plus a scalar field, justifying the three d.o.f.  coinciding with polarizations. 
It is worth stressing again that,  the above analysis includes the sub-cases $f(T,B)= f(-T,-B)=f(R)$ and $f(T,B)=f(T)+B=f(T)$ reported in literature. Clearly, the number of polarizations in $f(R)$ and $f(T)$ gravity are recovered. See \cite{AC, Hohmann1, Hohmann2, Hohmann3}.

Another motivation for the presence of  scalar mode is the symmetry breaking of the TT gauge due to the massive wave, that is, the mass of scalar field brakes the symmetry of the TT gauge. 
In GR the absence
of scalar, longitudinal, and vector modes implies that the response of detectors is governed entirely by the transverse-trace free modes. This fact is relevant to compare alternative theories with GR. In the case of  $f(T,B)$ gravity, it  is not possible to perform a gauge transformation on $E_{\mu\nu}$ that makes it traceless and completely spatial at the same time,  namely performing a  TT gauge. According to these considerations,   $f\left(T, B\right)$ gravity shows three polarizations: the two standard plus $(+)$ and cross $(\times)$ 2-helicity massless transverse tensor polarization modes and a 0-helicity massive scalar polarization mode $(s)$, resulting as a mixed state of longitudinal and breathing transverse polarizations. 
Being dynamically equivalent to $f(R)$ gravity, it is possible to show that, in the post-Newtonian limit, a Yukawa-like correction emerges in general \cite{Yukawa}. This correction can be considered to put upper bound on the graviton mass as discussed in \cite{Zakharov1,Zakharov2,Zakharov3}.
Being $f(R)$ gravity not excluded by observations \cite{GW17_1,GW17_2,GW17_3}, this could be a pathway to test also $f(T,B)$ gravity by a possible massive mode.

 An important remark is in order at this point.  Besides  perturbations around the Minkowski background, it is interesting to develop a similar analysis  around a  cosmological background.  This approach  results useful  to investigate  primordial gravitational waves. For example,  assuming a Friedmann-Robertson-Walker spatially flat metric as 
\begin{equation}
ds^2=dt^2-a^2(t)\delta_{ij}dx^{i}dx^{j}\ ,
\end{equation} 
with $a(t)$ the scale factor, we can perturb the tetrad $e^{a}_{\phantom{a}\mu}=\text{diag}(1,a(t),a(t),a(t))$  obtaining 
\begin{equation}\label{perturbed tetrad}
e^{a}_{\phantom{a}\mu}=\bar{e}^{a}_{\phantom{a}\mu}+E^{a}_{\phantom{a}\mu}\ ,
\end{equation}
where $\bar{e}^{a}_{\phantom{a}\mu}$ is the unperturbed part of the tetrad. See \cite{CLSX, NPS} for details. Then, inserting the above cosmological  metric into the field  Eqs. \eqref{EqFielTB}, we obtain the related  Friedmann equations. From  Eq. \eqref{perturbed tetrad}, it is possible  to derive the differential equations for  $E^{a}_{\mu}$ giving rise to  cosmological gravitational waves as solutions. This kind of analysis has been developed in  \cite{CDON} for $f(R)$.
In a forthcoming paper, cosmological gravitational waves for generalized TEGR theories will be discussed.

\begin{acknowledgements}
 SC acknowledges the support of  INFN ({\it iniziative specifiche} MOONLIGHT2 and QGSKY).
 This paper is based upon work from COST action CA15117 (CANTATA), supported by COST (European Cooperation in Science and Technology).
\end{acknowledgements}

\end{document}